\documentclass[a4paper,11pt]{article}
\usepackage{algorithm}
\usepackage{algorithmic}
\usepackage{amsmath}
\usepackage{fancyhdr,lastpage}
\usepackage{enumerate}
\usepackage{graphicx}
\usepackage{epstopdf}
\usepackage[round]{natbib}

\newcommand{\expv}[1]{\left \langle #1 \right \rangle}

\begin{document}
\title{Rotational diffusion of particles in turbulence}
\author{Colin R. Meyer and Evan A. Variano}
\maketitle
\begin{abstract}
Through laboratory measurements, we compare the rotation of spherical and ellipsoidal particles in homogeneous, isotropic turbulence. We find that the particles' angular velocity statistics are well described by an Ornstein-Uhlenbeck (OU) process.  This theoretical model predicts that the Lagrangian autocovariance of particles' angular velocity will decay exponentially.  We measure the autocovariance using stereoscopic particle image velocimetry (SPIV) applied to particles whose size is within the inertial subrange of the ambient turbulence.  The SPIV resolves the motion of points interior to the particles, from which we calculate the solid body rotation of the particles.  This provides us with the angular velocity time series for individual particles.  Through ensemble statistics, we determine the autocovariance of angular velocity and confirm that it matches the form predicted by an OU process.  We can further use the autocovariance curve to quantify the turbulent rotational diffusivity. 
\end{abstract}

\section{\label{Introduction}Introduction}
In this paper, we compare rotational dynamics of spherical and non-spherical particles suspended in  a turbulent flow. Rotation of particles is important in environmental and industrial processes, such as flocculation and papermaking, in which rotation can affect particle aggregation \citep{Koc1989}. In these and many other applications, particles are typically non-spherical.  This leads to close coupling of the rotational and translational motions and further complicates the mathematical treatment \citep{Mor2008b, Cli2005}.
 
For both spherical and non-spherical particles, rotation significantly effects the wake structure behind a particle and thus the forces coupling the particle and surrounding fluid \citep{Bat1967, Gia2009}.  These forces, in turn, set the particle rotation rate by causing angular acceleration \citep{Bag2002, Shi2005}.  This feedback can lead to complex behaviors, e.g. broken symmetry in quiescent fluid \citep{Jen2004}; rotation in a direction opposite to local vorticity during the transient approach to equilibrium in a steady shear flow \citep{Bag2002}; and spherical particles that do not rotate at half the local fluid vorticity in turbulent flow \citep{Mor2007, Mor2008a}.

Given the complexities of directly predicting particle rotation in turbulent flow, a stochastic approach is a useful alternative.  In analogy to a translational random walk, we will model the time-dependent evolution of a particles' angular velocity as a random walk in rotation space. A stochastic process that closely approximates the translational motion of fluid parcels in turbulence is the Ornstein-Uhlenbeck (OU) process. The properties of this process are described in the next section but, importantly, it describes a diffusion process that is stationary, Markovian, and Gaussian. This paper hypothesizes that the OU process can be applied to the angular velocity of spherical and ellipsoidal particles, resulting in rotational diffusion that is analogous to the translational diffusion. We test this hypothesis by measuring the Lagrangian autocovariance and probability density functions (PDFs) of particles' angular velocity. Understanding rotational diffusion will allow prediction of the dynamics of particle orientation in turbulent flows

\section{\label{Theory}Theory}
The Langevin equation is a stochastic differential equation that describes the Ornstein-Uhlenbeck (OU) stochastic process. The OU process is completely characterized by a Gaussian distribution and an autocovariance, both of which are statistically stationary \citep{Doo1942}.  This stochastic process is Markovian, meaning that the probability distribution of the next state is determined only by the probability distribution of the current state \citep{Nor1998}.  In the OU process, the autocovariance takes the specific form of a negative exponential, e.g. $R(\tau)=\sigma^2 e^{-\tau/T}$, where $\sigma^2$ is the variance of the Gaussian distribution, $\tau$ is the time--lag, and $T$ is the integral timescale.  The OU process has been shown to accurately describe the Lagrangian velocity of fluid parcels in turbulent flow \citep{Pop2000}.  One important implication of this is that OU process and Langevin equation can be used to accurately model the correlated random walk that describes the diffusion of a cloud of particles.  The time-dependent diffusivity is given by $\Gamma(t)=\int_{0}^{t}{R(\tau)d\tau}$, and can be used to apply Fick's law for mass flux in a turbulent flow.

We hypothesize that the OU process and Langevin equation are also a good model for the rotation of large particles suspended in a turbulent flow.   Specifically, we hypothesize that the Lagrangian timeseries of angular velocities will follow an OU process.  We can test this hypothesis by determining whether the Lagrangian angular velocity autocovariance decays exponentially as 
\begin{equation}
R_{\Omega_i}(\tau)=\alpha_i^2 e^{-\frac{\tau}{T_{\Omega_i}}},
\label{AutoC}
\end{equation}
defining $T_{\Omega_i}$ as the integral timescale of Lagrangian angular velocity, and $\alpha_i^2$ as the angular velocity variance. We also check whether the PDF of particles' angular velocity is stationary and Gaussian. In \S \ref{Results} we see that observations match these predictions quite well.  As a result, we can find the time dependent rotational turbulent diffusivity \citep{Tay1921} as
\begin{equation}
\Gamma_{\Omega_i} (t) =\int_{0}^{t}{R_{\Omega_i} (\tau) d\tau}=\alpha_i^2 T_{\Omega_i} \left(1-e^{-\frac{t}{T_{\Omega_i}} }\right).
\label{DiffrotE}
\end{equation}
The rotational diffusity allows us to calculate the characteristic time for a complete rotation cycle. 

\section{\label{Methods}Methods}
We study particle motion in homogeneous, isotropic turbulence generated in a laboratory facility.  This is a symmetrically stirred rectangular tank driven by a planar forcing element on each end. The symmetry contributes to large scale isotropy, as does the use of two randomly actuated synthetic jet arrays (RASJAs) for the forcing elements \citep{Var2008}. Each RASJA is a planar array of 64 synthetic jets. The jets fire in a stochastic pattern that maximizes the shear production of turbulence, similar to an active grid wind tunnel but without mean flow. The RASJAs allow the tank-scale secondary circulations to be essentially eliminated, which increases the residence time of particles in the image area helping us measure Lagrangian particle trajectories. Two screens are placed between the pumps and the center of the tank, which keep the particles contained in the central measurement region. The flow properties of the RASJAs without screens are discussed extensively in \citet{Bel2012}. The turbulent statistics for the RASJAs with screens and without particles can be found in table \ref{fparams}. Although the screens decrease the turbulent kinetic energy, they do not affect the homogeneity or isotropy of the flow.

We measure the turbulent statistics of the fluid--phase (without particles) using 2D particle image velocimetry (2DPIV). In the center region of the tank, the Reynolds number (defined with respect to the Taylor microscale) is $\mbox{Re}_{\lambda} = 110$. Single-point turbulent statistics are homogeneous and isotropic throughout a large center region. In previous experiments we found that this region is much larger than the turbulent integral length scale; therefore, the measurements are not affected by spatial heterogeneity. 

\begin{table}
\begin{center}
\scriptsize
\begin{tabular}{llcc}
\hline
Transverse velocity rms, & $u_{rms}$ (m s$^{-1}$) & $\approx0.012$ & -\\
Vertical velocity rms, &  $v_{rms}$ (m s$^{-1}$) & $0.012$ & [0.012\ 0.012]\\
Longitudinal velocity rms, &  $w_{rms}$ (m s$^{-1}$) & $0.013$ & [0.013\ 0.013] \\
Turbulent kinetic energy, &  k ($=\frac{1}{2}(\sum_1^3{u_{i_{rms}}^2}$)) (m$^2$ s$^{-2}$) & $2.3\times10^{-4}$ & [$2.3\times10^{-4}$\ $2.3\times10^{-4}$] \\
Taylor microscale, &	$\lambda_z$ (m)  & $8.3\times10^{-3}$ & [6.2$\times10^{-3}$\ 10$\times10^{-3}$]\\
Integral length-scale, &  $\Lambda_z$ (m)  & $57\times10^{-3}$ & [$57\times10^{-3}$\ $58\times10^{-3}$]\\
Eddy turnover time, &  $T$ ($=\Lambda_z/w_{rms}$) (s)        & $4.3$ & [4.3\ 4.3]\\
Kinematic viscosity, &  $\nu$ (m$^2$ s$^{-1}$)       & $1.0\times10^{-6}$ & - \\
Reynolds number (Taylor), &  $\mbox{Re}_{\lambda}$ ($=w_{rms}\lambda_z/\nu$)   & 110 & [81\ 140]\\
Reynolds number (Integral), &  $\mbox{Re}_{\Lambda}$ ($=w_{rms}\Lambda_z/\nu$)                 & 760\ & [750\ 760]\\
Turbulent dissapation rate, &  $\epsilon$ $(=15 \nu [w_{rms}^2 / \lambda_z^2]$) (m$^2$ s$^{-3}$)       & $3.8\times10^{-5}$ & [$1.9\times10^{-5}$\ $5.8\times10^{-5}$]\\
Kolmogorov length-scale, &  $\eta$ ($=(\nu^3/\epsilon)^{1/4}$) (m)   & $0.40\times10^{-3}$ & [$0.40\times10^{-3}$\ $0.40\times10^{-3}$]\\
Kolmogorov time-scale, &  $\tau_{\eta}$ ($=(\nu/\epsilon)^{1/2}$) (s)    & $0.16$ & [0.16\ 0.17]\\
\hline
\end{tabular}
\end{center}
\caption{Turbulent statistics of fluid-phase without particles in experimental facility with 95\% confidence intervals. Both $\lambda_z$ and $\Lambda_z$ are calculated from the longitudinal autocovariance of fluid--phase velocity. The introduction of particles significantly reduces the turbulent kinetic energy; spheres reduce $k$ by 15\% and ellipsoids reduce $k$ by 3\% \citep{Bel2012}.} 
\label{fparams}
\end{table}

We place hydrogel spheres or ellipsoids (made of an agarose solution of 4 g/L) into the stirred tank. The spherical particles have diameter of $d_s$ = 8 mm and the ellipsoidal particles have polar and equatorial axes of $l_e$ = 16 mm and $d_e$ = 8 mm respectively. These dimensions are such that they match the Taylor length scale (from table \ref{fparams}, $\lambda_z=8.3$ mm) and their relaxation time scale (3.64 s for spheres and 5.46 s for ellipsoids, calculated using the major axis) is close to the eddy turnover time of the ambient turbulence ( $T=4.3$ s). The particle Reynolds numbers are much greater than 1 and are 22 and 63 for spheres and ellipsoids respectively \citep{Bel2012b}. The particles are near-neutrally buoyant (1007 kg m$^{-3}$), thus they stay suspended during experiments. They are refractively matched to water and nearly transparent.  These optical properties are essential, as they allow us to pass a laser sheet through the particles and measure the motion of interior points via stereoscopic particle image velocimetry (SPIV).  Interior points are made visible by reflective tracers embedded in the hydrogel. 

\begin{table}
\begin{center}
\begin{tabular}{lc}
\hline
  $d_s$ (m)         &     $8\times10^{-3}$\\
  $d_e$ (m)         &     $16\times10^{-3}$\\
  $l_e$ (m)         &     $8\times10^{-3}$\\
  $\rho_p$ (kg m$^{-3}$)         &     $1007$\\
\hline
  \end{tabular}
\end{center}
\caption{Quantitative description of hydrogel particles.} 
\label{pparams}
\end{table}

Using a SPIV system (commercially provided by LaVision, Inc.), we find the flow field for the fluid--phase turbulence as well as the velocity vectors inside the particles. Our temporal resolution is 14.773 Hz, which is faster than the particle relaxation timescale \citep[see][]{Bel2012} and our spatial resolution is 1.38 mm, which is smaller than the particle size. We measure velocity vectors inside the solid particles by tracking clusters of the embedded tracers via standard SPIV. The resulting three-component velocity measurements along a 2D plane inside each particle allows us to calculate the particle angular velocity, $\mathbf{\Omega}\ = [\Omega_{x}, \Omega_{y}, \Omega_{z}]$.  We do so by using the solid body rotation equation:
\begin{equation}
\mathbf{V}_{L} = \mathbf{V}_{M} + \mathbf{\Omega} \times \mathbf{R}_{LM},
\label{solidbody}
\end{equation}
where $\mathbf{V}_L$ is the velocity at any interior point $L$, $\mathbf{V}_M$ is the velocity at a second interior point $M$, $\mathbf{\Omega}$ is the rotation about the center of mass, and $\mathbf{R}_{LM}$ is the distance vector between points $L$ and $M$. Due to the fact that we measure three components of velocity at each point (e.g. $\mathbf{V}_L = [V_{Lx}, V_{Ly}, V_{Lz}]$) and all points are coplanar (i.e. $\mathbf{R}_{LM} = [R_{LMx}, R_{LMy}, 0]$), this system of equations is over-determined in the $z$-coordinate and under-determined in the $x$- and $y$-coordinates. By including a third interior point ($N$), equation \eqref{solidbody} can be rewritten twice, once in terms of $L$ and $N$ as well as once in terms of $M$ and $N$. Using equations \eqref{solidbody} and the two other permutations, all three components of rotation can be determined. This method gives a single value for $\Omega_x$ as well as $\Omega_y$ and four estimates of $\Omega_z$. We average the four estimates of $\Omega_z$, therefore our result for $\Omega_z$ contains less measurement noise than the other two components.  Since there are multiple velocity points inside of the particle, we calculate $\mathbf{\Omega}$ once for each possible triplet, $V_L$, $V_N$ and $V_M$. Compiling all of the measurements of $\mathbf{\Omega}$ for a single particle gives a distribution for each component. The median value for each component is selected as the best estimate of particle angular velocity \citep{Bel2012}. 

We measure $\mathbf{\Omega}$ for a particle as it travels along a trajectory through the turbulent flow. These trajectories range in length from 8 to 119 SPIV frames. Collecting $\mathbf{\Omega}$ for all frames gives us a time series of particle rotation. We then compute the autocovariance for each component of $\Omega_i(t)$ from its definition:
\begin{equation}
R_{\Omega_i} (\tau) = \langle \Omega_i(t+\tau) \Omega_i(t) \rangle.
\label{autocovfun}
\end{equation}
Here $\expv{\cdot}$ signifies an ensemble average across all particle trajectories. 

We collect sufficient trajectories to calculate statistically converged data in the first and second moments of the distribution of $\Omega_i$.  That is, adding more data would decrease the uncertainty on the moments, but not change their values. Our data show convergence for the mean and variance of $\mathbf{\Omega}$ after 300 independent time series.  In total, we collected 407 ellipsoids trajectories and 572 sphere trajectories.  With this data, we were able to also compute a fourth standardized moment (kurtosis) that converged for $\Omega_z$ (the low--noise component) but not for the other two components. \citet{Sny1971} predict that 700 independent time series are needed to reduce the error below 10\% in the measurement of the autocorrelation of translational velocity statistics. The value of 700 independent time series is slightly larger than value of 300, which we measure. This makes sense because we measure the autocovariance instead of the autocorrelation, and the former converges with fewer samples. 

The definition of $\mathbf{\Omega}$ used here is the angular velocity of a particle about its center of mass, with the three rotation components referenced to a set of coordinate axes that are fixed in the laboratory frame.  It would also be possible, though more complicated experimentally, to decompose the rotation into the components relative to a set of coordinate axes that move with each particle.  The two definitions will likely not give identical results, and both would be interesting to know.  For example, when studying organisms, the rotation relative to the local axes would be useful in quantifying changes to motile diffusion and drag, and the rotation relative to the laboratory axes would be useful in determining the statistics of the organism's sensory search of 3D space.  Herein, we focus only on the rotation relative to the lab coordinates, and evaluate this rotation while following each particle on its translational random walk.  This is the exact meaning of our definition of ``Lagrangian angular velocity," and we caution readers to be aware of the other possible definition.

Because our data is transformed several times between the raw measurement (tracer locations in a plane) to the final measurement (particle angular velocity vector), it is nontrivial to propogate error to find the uncertainty in our measurements.  Thus we use a Monte Carlo (MC) simulation to evaluate the uncertainty of $\mathbf{\Omega}$. We begin by constructing synthetic angular velocity trajectories using an OU process \citep{Gil1996}.  We then construct corresponding planar velocity vector fields of the type which we measure using SPIV.  To these vector fields data, we add random noise that is normally distributed with zero mean. Finally, both the noisy and noise-free simulated vector fields are passed through the rotation algorithm to calculate $\mathbf{\Omega}$.  By comparing the results from the noise-free and noisy simulated measurements, we can assess the effects of measurement noise on the statistics of $\mathbf{\Omega}$; the results are discussed in the next section.

\section{\label{Results}Results}
\begin{figure}[!ht]
\begin{minipage}[b]{0.48\linewidth}
\centering
\includegraphics[scale=0.55]{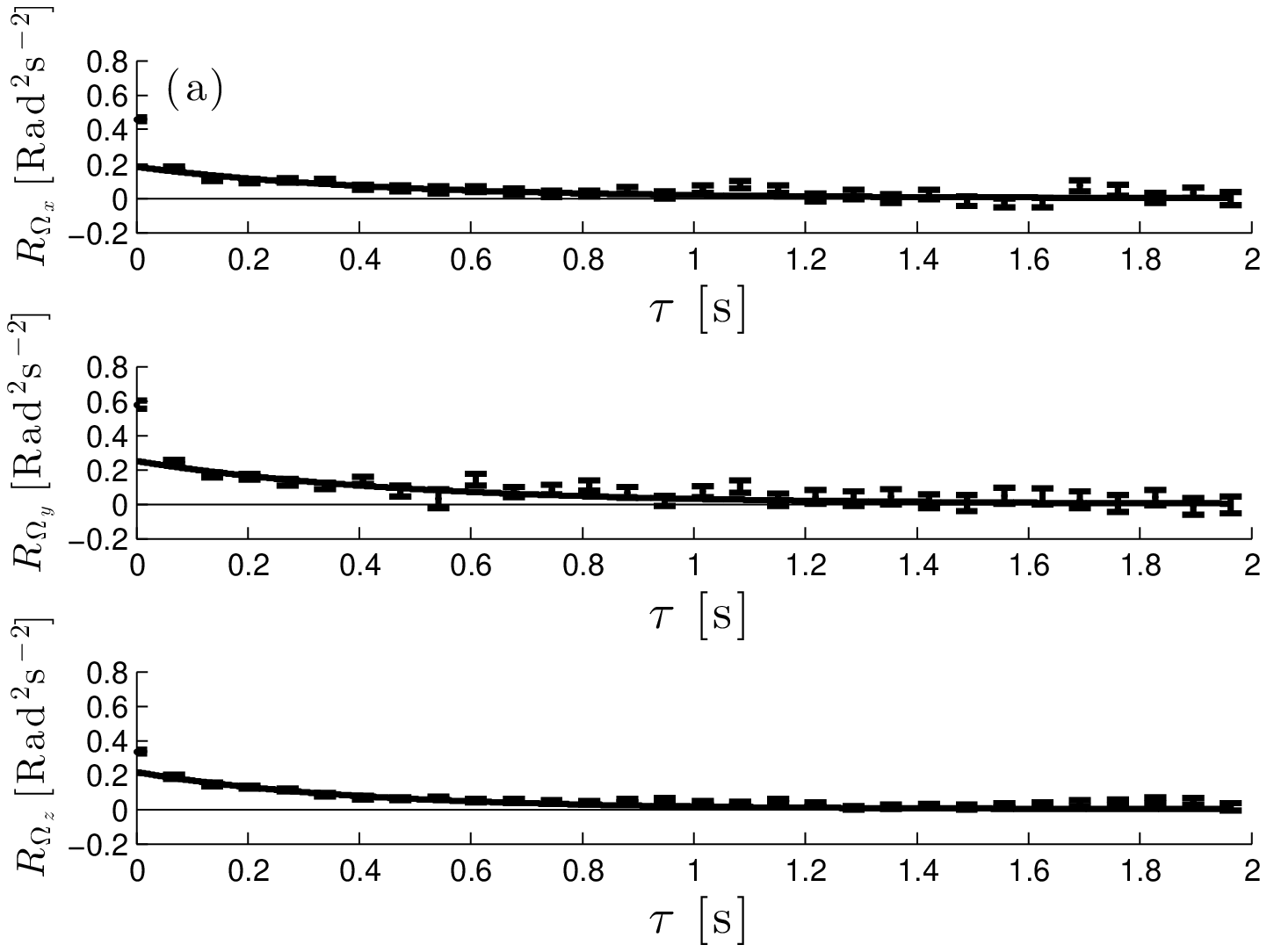}
\end{minipage}
\begin{minipage}[b]{0.48\linewidth}
\centering\
\includegraphics[scale=0.55]{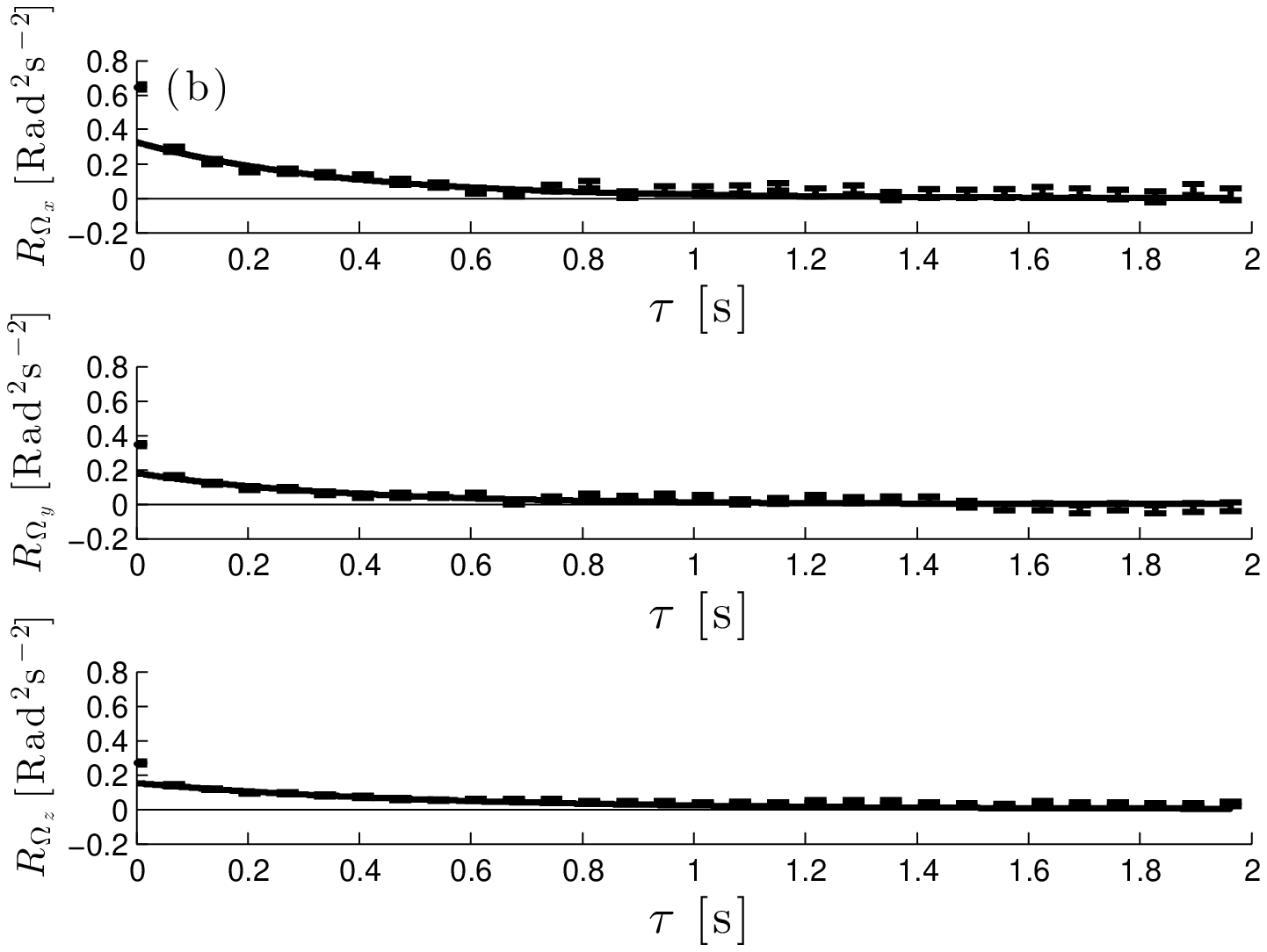}
\end{minipage}
\caption{Autocovariance of rotation for (a) spheres. (b) ellipsoids. Top panel is $\Omega_x$, middle panel is $\Omega_y$, and bottom panel is $\Omega_z$. Points are experimental data. Solid line is exponential fit. Error bars are standard error.}
\label{autocov}
\end{figure}

The experimental autocovariance curves for the spherical and ellipsoidal particles are shown in figures \ref{autocov}a and \ref{autocov}b. Equation \eqref{AutoC} is fit to the data with free parameters $\alpha_i^2$ and $T_{\Omega_i}$, reported in table \ref{fitparams}. The characteristic timescale ($T_{\Omega_i}$) of exponential decay found in these fits is the Lagrangian integral timescale of particle rotation, and $\alpha_i^2$ is the variance of particle angular velocity . 

\begin{table}
\begin{center}
\large
\begin{tabular}{lcccc}
\hline
& Spheres & 95\% CI & Ellipsoids & 95\% CI \\
\hline
  $\alpha_x$ (Rad s$^{-1}$) & 0.43 & [0.39\ 0.47]  & 0.57 & [0.54\ 0.61]\\
  $\alpha_y$ (Rad s$^{-1}$) & 0.50 & [0.40\ 0.60]  & 0.42 & [0.39\ 0.46]\\
  $\alpha_z$ (Rad s$^{-1}$) & 0.47 & [0.44\ 0.49]  & 0.39 & [0.38\ 0.41]\\
  $T_{\Omega_x}$ (s) & 0.43 & [0.29\ 0.57]   & 0.37 & [0.30\ 0.43]\\
  $T_{\Omega_y}$ (s) & 0.48 & [0.15\ 0.81]   & 0.37 & [0.28\ 0.45]\\
  $T_{\Omega_z}$ (s) & 0.39 & [0.33\ 0.45]   & 0.52 & [0.45\ 0.59]\\
\hline
\end{tabular}
\end{center}
\caption{Parameters from the exponential fit: noise-free standard deviation ($\alpha_i$) and Lagrangian integral timescale ($T_{\Omega_i}$).} 
\label{fitparams} 
\end{table}

\begin{figure}[!ht]
\centering
\includegraphics[scale=0.72]{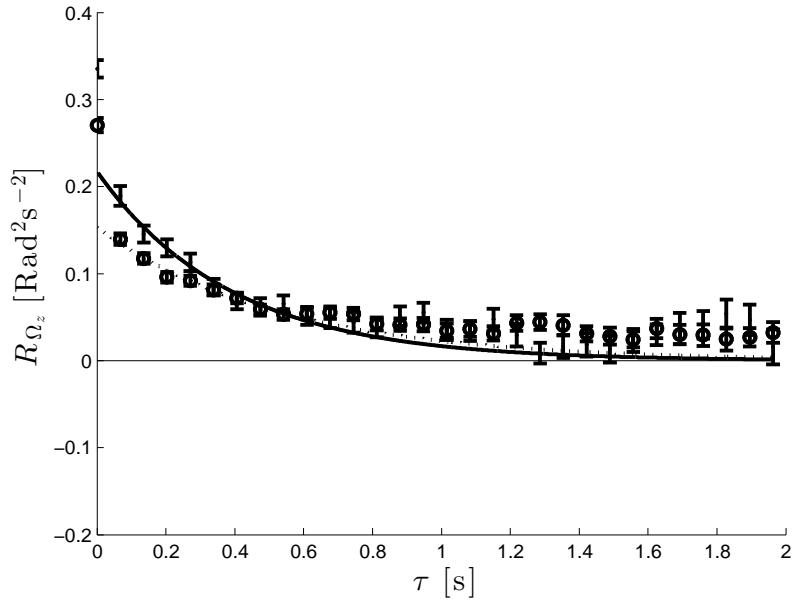}
\caption{Autocovariance of rotation about the z-axis ($\Omega_z$) for spheres (points and solid line) and ellipsoids (circles and dashed line). For both plots, error bars are standard error. Due to a large number of short trajectories, fit is closer for small lag time, $\tau$.}
\label{SandEautocov}
\end{figure}

Evident in figures \ref{autocov} and \ref{SandEautocov} is a considerable deviation between data and model at zero lag ($\tau = 0$) due to measurement noise. If this measurement noise is uncorrelated with itself over time, then the noise will be present only in the first point on each curve. Until lags greater than about a second and a half, this appears to be the case, because the data and exponential curve fit agree well. For larger lags, because there are many more short trajectories than long trajectories in our dataset, the fit and data are more discrepant (see figure \ref{SandEautocov}).  The exponential fit allows us to estimate the noise-free variance at $R_{\Omega_i}(\tau = 0)$, which is equal to $\alpha_i^2$.  
 
We approximate the magnitude of measurement noise as the difference between the variance measured as the second moment of the angular velocity PDF ($\sigma_i^2$) and the noise-free variance measured from the exponential fit to autocovariance ($\alpha_i^2$). From these values we define a signal-to-noise ratio as 
\begin{equation}
\mbox{SNR}_i=\sqrt{\frac{\alpha_i^2}{\sigma_i^2-\alpha_i^2}}.
\label{SNReq}
\end{equation} 
The SNR, measured variance, and noise-free variance are all reported in table \ref{rparams}. The SNR is an essential input into our Monte Carlo analysis; we set the simulated measurement noise in the Monte Carlo simulation so that its SNR matches that which we observe in the measurements.  We then use the MC results to infer how measurement noise affects statistics other than the angular velocity variance.   

\begin{table}
\begin{center}
\tiny
\begin{tabular}{lcccccccc}
\hline
& \multicolumn{4}{c}{Stereoscopic Particle Image Velocimetry (SPIV)} & \multicolumn{4}{c}{Monte Carlo (MC) Simulation} \\
& Spheres & 95\% CI & Ellipsoids & 95\% CI & Initial & 95\% CI & Final & 95\% CI \\
\hline
	SNR$_x$  & 0.81 & [0.72\ 0.91] & 1.0 & [0.92\ 1.1] & - & - & 0.85 & [0.89\ 0.95]\\
	SNR$_y$  & 0.88 & [0.66\ 1.1] & 1.0 & [0.92\ 1.2] & - & - & 0.87 & [0.90\ 0.97]\\
	SNR$_z$  & 1.4 & [1.2\ 1.5] & 1.1 & [1.1\ 1.2] & - & - & 1.3 & [1.3\ 1.4]\\
  $\expv{\Omega_x}$ (Rad s$^{-1}$) & -0.012 & [-0.036\ 0.010]  & -0.024 & [-0.050\ -0.00043] & 0.033 & [0.014\ 0.052] & 0.018 & [-0.014\ 0.047]\\
  $\expv{\Omega_y}$ (Rad s$^{-1}$) &  -0.029 & [-0.054\ 0.0072]  & -0.052 & [-0.077\ -0.032] & 0.033 & [0.014\ 0.052] & 0.032 & [0.0023\ 0.061]\\
  $\expv{\Omega_z}$ (Rad s$^{-1}$) & 0.021 & [0.0042\ 0.036]  &  0.011 & [-0.0041\ 0.025] & 0.033 & [0.014\ 0.052] & 0.031 & [0.0063\ 0.058]\\
  $\sigma_x$ (Rad s$^{-1}$) & 0.68 & [0.66\ 0.69]  & 0.80 & [0.79\ 0.82] & 1.0 & [0.99\ 1.0] & 1.5 & [1.5\ 1.5] \\
  $\sigma_y$ (Rad s$^{-1}$) & 0.76 & [0.73\ 0.79]  & 0.59 & [0.57\ 0.61] & 1.0 & [0.99\ 1.0] & 1.5 & [1.4\ 1.5]\\
  $\sigma_z$ (Rad s$^{-1}$) & 0.58 & [0.56\ 0.60]  & 0.52 & [0.50\ 0.53] & 1.0 & [0.99\ 1.0] & 1.3 & [1.2\ 1.3]\\
  $\alpha_x$ (Rad s$^{-1}$) & 0.43 & [0.39\ 0.47]  & 0.57 & [0.54\ 0.61] & 1.0 & [0.99\ 1.0] & 0.97 & [0.94\ 1.0]\\
  $\alpha_y$ (Rad s$^{-1}$) & 0.50 & [0.40\ 0.60]  & 0.42 & [0.39\ 0.46] & 1.0 & [0.99\ 1.0] & 0.98 & [0.95\ 1.0]\\
  $\alpha_z$ (Rad s$^{-1}$) & 0.47 & [0.44\ 0.49]  & 0.39 & [0.38\ 0.41] & 1.0 & [0.99\ 1.0] & 1.01 & [0.99\ 1.0]\\
  $K_x$  & $\approx$2.7 & - & $\approx$2.7 & - & 3.0 & [2.9\ 3.0] & 8.2 & [7.3\ 9.2]\\
  $K_y$  & $\approx$5.1 & - & $\approx$2.9 & - & 3.0 & [2.9\ 3.0] & 7.7 & [6.8\ 8.7]\\
  $K_z$  & 5.7 & [4.9\ 6.7]  & 5.3 & [4.8\ 5.9] & 3.0 & [2.9\ 3.0] & 6.6 & [6.0\ 7.2]\\
\hline
\end{tabular}
\end{center}
\caption{Moments of angular velocity PDF and Monte Carlo simulation PDF. Here $\sigma_i$ is the standard deviation of the PDF of $\Omega_i$ and $K_i$ is the kurtosis of the PDF of $\Omega_i$. The SNR is calculated by measuring the noise magnitude as the difference between the directly measured standard devation ($\sigma_i$) and the noise-free standard devation ($\alpha_i$) obtained from the exponential fit. } 
\label{rparams} 
\end{table}

The rotation statistics measured from our data are given in table \ref{rparams}, along with corresponding MC results.  Due to the isotropy of the turbulence in the tank (see table \ref{fparams}), the rotation of the particles should also be isotropic. The fitted parameters on the exponential model (table \ref{fitparams}) indicate that this is true, to within measurement uncertainty.  As expected, the effect of measurement noise is greater for rotation about the $x$- and $y$- axes than the $z$-axis.  This is because in the algorithm to compute rotation, $\Omega_z$ is the overdetermined component and therefore contains more data, increasing accuracy (see \S \ref{Methods}). This anisotropic noise effect can be seen clearly in the Monte Carlo results (table \ref{rparams});  when statistically isotropic rotation (seen in the column marked ``Initial'') is measured with our anisotropic imaging routine, the resulting PDF has anisotropic moments.  

Noise-free variance can be easily calculated as discussed above, but no such method is available for kurtosis.  Thus to understand kurtosis we use the MC analysis as a guide. MC results show that our measurement method gives kurtosis values that are strongly upward-biased.  Thus a distribution of rotation values that is truly Gaussian ($K=3$) will appear super-Gaussian in our measurements.  This bias is present in all coordinate directions, though it is stronger in $x$ and $y$ than in $z$.  The biased measurements of kurtosis computed from our experiments are seen in table \ref{rparams}; only the $z$-component is statistically converged, thus we do not report confidence intervals for the others.  The measured values of $K_z$ are close to the MC predictions for a biased measurement of a Gaussian distribution.  In fact, they are slightly smaller than these predictions.  We conclude from this that the true distribution of angular velocities in our experiment is near Gaussian, and may even be slightly sub-Gaussian ($K<3$).

\section{\label{Discussion}Discussion }
The hypothesis of this paper is that the rotation of large particles suspended in turbulence satisfies an OU process. As mentioned in \S \ref{Theory}, the three characteristics of this stochastic process are: exponentially decaying autocovariance, Gaussian distribution, and stationarity. Other than the noise at zero-lag, figures \ref{autocov}a and \ref{autocov}b show that the exponential decay model is quite effective for particle rotation.  Based on the discussion of kurtosis above, we conclude that the distribution of angular velocities is likely close to Gaussian.  We can assume that the rotation process is stationary because the driving flow is both stationary and spatially homogeneous. With these three observations as support, we conclude that the OU process is an acceptable model for the Lagrangian angular velocity of large particles.

\begin{figure}
\begin{minipage}[b]{0.48\linewidth}
\centering\
\includegraphics[scale=0.55]{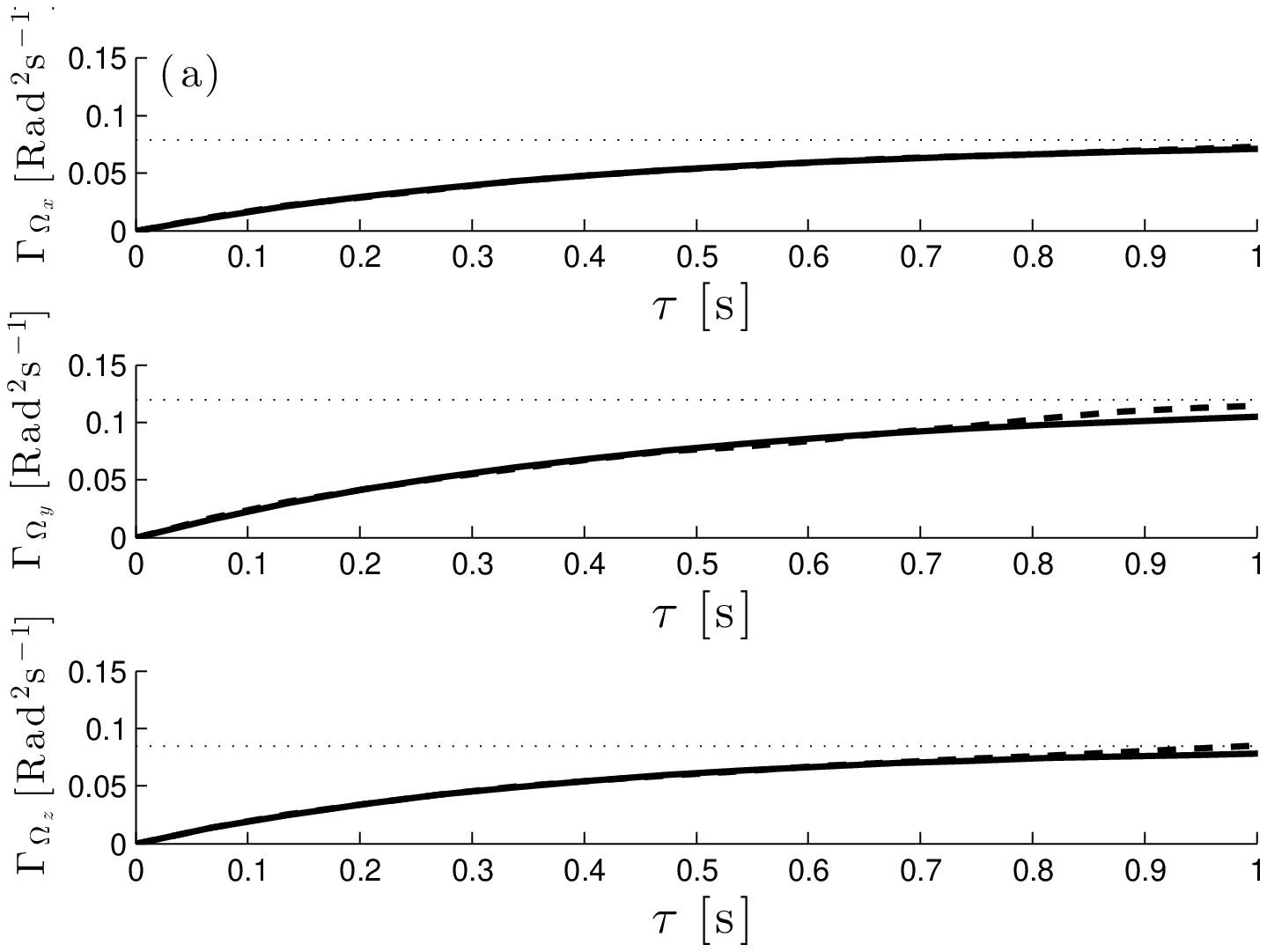}
\end{minipage}
\begin{minipage}[b]{0.48\linewidth}
\centering\
\includegraphics[scale=0.55]{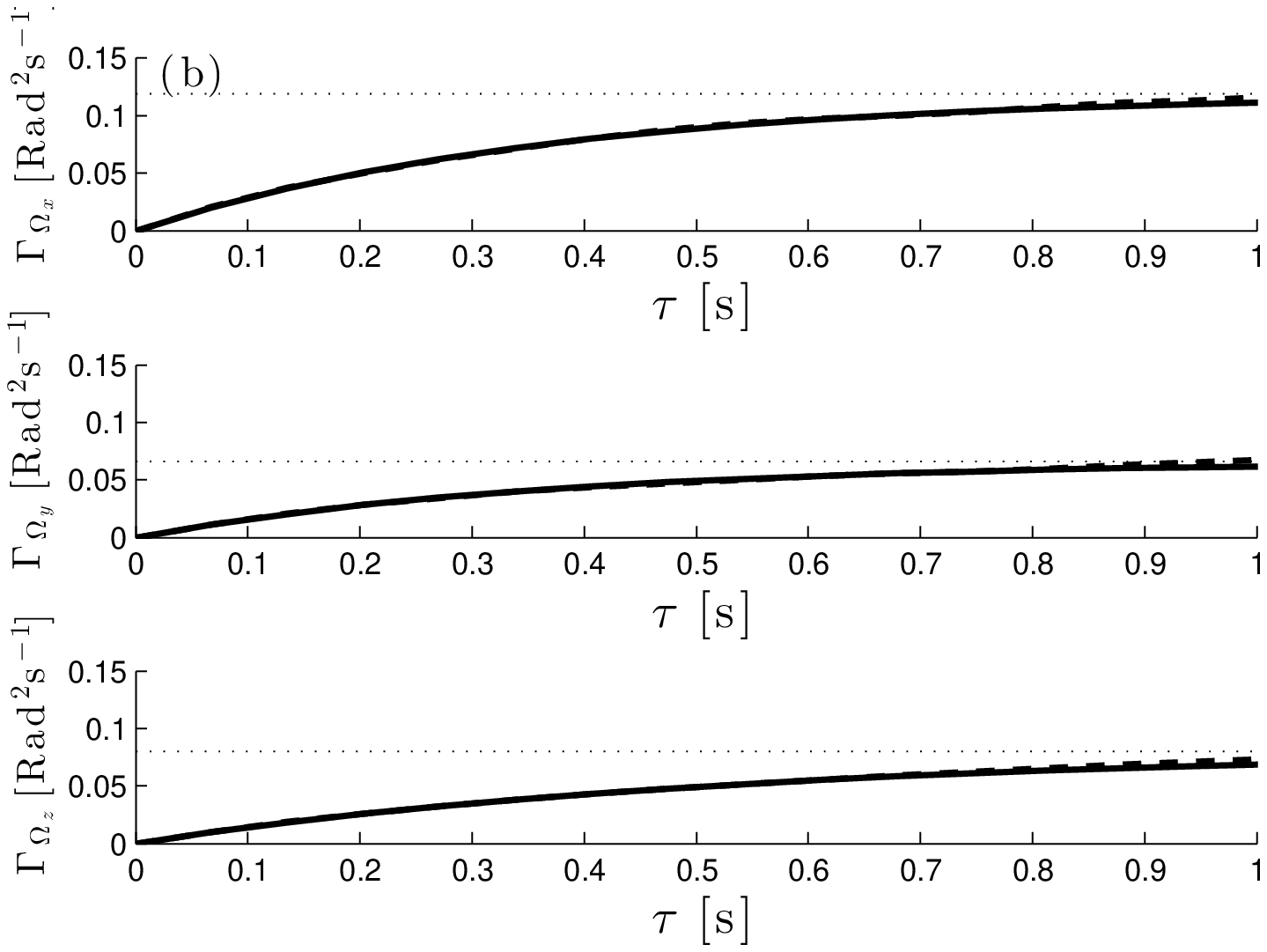}
\end{minipage}
\caption{Time dependent rotational diffusivity. (a) Spheres and (b) Ellipsoids. Dashed line is experimental data (noise removed), solid line is equation \eqref{DiffrotE}, and dotted line is the Fickian asymptote.}
\label{rotdiff}
\end{figure}

We can use the OU process to determine the time-dependent rotational diffusivity of large particles in turbulent flow, $\Gamma_{\Omega_i}(t)$.  This is given by equation \eqref{DiffrotE}, and we have obtained $\alpha_i^2$ and $T_{\Omega_i}$ from the exponential fit to our measurements. Figures \ref{rotdiff}a and \ref{rotdiff}b show the time-varying diffusivity predicted in this way.  When the exponential term diminishes to a negligible value, the $\Gamma_{\Omega_i}(\tau)$ tends toward a constant value: the Fickian asymptote. The value of the asymptote is calculated as $\hat{\Gamma}_{\Omega_i}=\alpha_i^2 T_{\Omega_i}$, and reported in table \ref{rotdiffparams}.  From these results we see that (for the flow-particle combination studied here), particles will diffusively complete one entire rotation cycle in an average of roughly 6 minutes ($T_{2\pi} = ((2\pi)^2/\hat{\Gamma}_{\Omega_i}\approx360$ s)).

\begin{table}
\begin{center}
\scriptsize 
\begin{tabular}{lcccccccc}
\hline
& \multicolumn{4}{c}{Stereoscopic Particle Image Velocimetry (SPIV)} & \multicolumn{4}{c}{Monte Carlo (MC) Simulation} \\
& Spheres & 95\% CI & Ellipsoids & 95\% CI & Initial & 95\% CI & Final & 95\% CI \\
\hline
  $\alpha_x$ (Rad s$^{-1}$) & 0.43 & [0.39\ 0.47]  & 0.57 & [0.54\ 0.61] & 1.0 & [0.99\ 1.0] & 0.97 & [0.94\ 1.0]\\
  $\alpha_y$ (Rad s$^{-1}$) & 0.50 & [0.40\ 0.60]  & 0.42 & [0.39\ 0.46] & 1.0 & [0.99\ 1.0] & 0.98 & [0.95\ 1.0]\\
  $\alpha_z$ (Rad s$^{-1}$) & 0.47 & [0.44\ 0.49]  & 0.39 & [0.38\ 0.41]] & 1.0 & [0.99\ 1.0] & 1.01 & [0.99\ 1.0]\\
  $T_{L_x}$ (s) & 0.43 & [0.29\ 0.57]   & 0.37 & [0.30\ 0.43] & 0.13 & [0.13\ 0.14] & 0.13 & [0.12\ 0.14]\\
  $T_{L_y}$ (s) & 0.48 & [0.15\ 0.81]   & 0.37 & [0.28\ 0.45] & 0.13 & [0.13\ 0.14] & 0.13 & [0.12\ 0.14]\\
  $T_{L_z}$ (s) & 0.39 & [0.33\ 0.45]   & 0.52 & [0.45\ 0.59] & 0.13 & [0.13\ 0.14] & 0.13 & [0.12\ 0.14]\\
  $\hat{\Gamma}_{\Omega_x}$ (Rad$^2$ s$^{-1}$) & 0.079 & [0.048\ 0.11]   & 0.12 & [0.093\ 0.15] & 0.14 & [0.13\ 0.15] & 0.12 & [0.11\ 0.13]\\
  $\hat{\Gamma}_{\Omega_y}$ (Rad$^2$ s$^{-1}$) & 0.12 & [0.025\ 0.21]   & 0.066 & [0.048\ 0.084] & 0.14 & [0.13\ 0.15] & 0.12 & [0.11\ 0.13]\\
  $\hat{\Gamma}_{\Omega_z}$ (Rad$^2$ s$^{-1}$) & 0.086 & [0.069\ 0.10]   & 0.080 & [0.068\ 0.092] & 0.14 & [0.13\ 0.15] & 0.13 & [0.12\ 0.14]\\
\hline
\end{tabular}
\end{center}
\caption{Calculation of rotational diffusivity, $\hat{\Gamma}_{\Omega_i}$. Comparing the laboratory measurements to MC simulation show that measurement noise has a slight effect on the calculation of the rotational diffusivity. However, the rotational diffusivity is anisotropic, compounded from the anisotropy in $\alpha_i$ and $T_{\Omega_i}$} 
\label{rotdiffparams} 
\end{table}

Particle shape appears to have only a minor effect on rotation statistics.  The Lagrangian integral timescales of angular velocity are statistically identical for ellipsoidal and spherical particles (table \ref{fitparams} or \ref{rotdiffparams}).  Due to measurement uncertainty, it is not as conclusive whether the variance terms are statistically similar or different between the two particle shapes. The work of \citet{Bel2012} suggests that the rotation of particles is dominated by the large-scales of fluid motion, similar to translational motion \citep{Kom1974}. These results imply that the effect of shape is minor.

\section{\label{Conclusion}Conclusion}
For spherical and ellipsoidal particles suspended in homogeneous, isotropic turbulence, we have shown that the particle angular velocity can be modeled as an OU process.  This conclusion follows from our laboratory measurements showing that the Lagrangian autocovariance of particles' angular velocity decays exponentially (figure \ref{autocov}). Our measurements also show that the particle rotation PDF is close to Gaussian, a conclusion that requires the use of a Monte Carlo simulation to evaluate the propagation of measurement noise through our analysis.  The results of this demonstrates that certain statistics, e.g. the kurtosis, were affected more than others, e.g. the rotational diffusivity. Comparing particles of two different shapes, we find that ellipticity does not strongly influence rotational diffusion. Both ellipsoids and spheres exhibit similar integral timescales of rotation and angular velocity variances. This suggests that the rotational diffusion of large particles in turbulent flow can be approximated well without detailed considerations of particle shape.  

We would like to thank Gabriele Bellani and Margaret Byron for support in making the laboratory measurements, Matt Ritter and Audric Collignon for early development of the measurement technique, and Jason Lepore, Matt Ford, and Yoram Rubin for insightful discussions. This research was conducted through undergraduate coursework supported by the department of Civil and Environmental Engineering at the University of California, Berkeley.

\bibliography{EFM}

\begin{thebibliography}{20}
\providecommand{\natexlab}[1]{#1}
\providecommand{\url}[1]{\texttt{#1}}
\expandafter\ifx\csname urlstyle\endcsname\relax
  \providecommand{\doi}[1]{doi: #1}\else
  \providecommand{\doi}{doi: \begingroup \urlstyle{rm}\Url}\fi

\bibitem[Bagchi and Balachandar(2002)]{Bag2002}
P.~Bagchi and S.~Balachandar.
\newblock Effect of free rotation on the motion of a solid sphere in linear
  shear flow at moderate re.
\newblock \emph{Phys. Fluids}, 14\penalty0 (8):\penalty0 2719--2737, 2002.

\bibitem[Batchelor(1967)]{Bat1967}
G.~K. Batchelor.
\newblock \emph{An Introduction to Fluid Dynamics}.
\newblock Cambridge University Press, 1967.

\bibitem[Bellani and Variano(2012)]{Bel2012b}
G.~Bellani and E.~A. Variano.
\newblock Slip-velocity and drag of large neutrally-buoyant particles in
  turbulent flows.
\newblock \emph{Accepted New J. Phys.}, 2012.

\bibitem[Bellani et~al.(2012)Bellani, Byron, Collignon, Meyer, and
  Variano]{Bel2012}
G.~Bellani, M.~L. Byron, A.~G. Collignon, C.~R. Meyer, and E.~A. Variano.
\newblock Shape effects on turbulent modulation by large nearly neutrally
  buoyant particles.
\newblock \emph{J. Fluid Mech.}, 45:\penalty0 35--97, 2012.

\bibitem[Clift et~al.(2005)Clift, Grace, and Weber]{Cli2005}
R.~Clift, J.R. Grace, and M.E. Weber.
\newblock \emph{Bubbles, Drops, and Particles}.
\newblock Dover Publications, 2005.

\bibitem[Doob(1942)]{Doo1942}
J.~L. Doob.
\newblock The brownian movement and stochastic equations.
\newblock \emph{Ann. Math.}, 43:\penalty0 351--369, 1942.
\newblock Reprinted in Selected Papers on Noise and Stochastic Processes.

\bibitem[Giacobello et~al.(2009)Giacobello, Ooi, and Balachandar]{Gia2009}
M.~Giacobello, A.~Ooi, and S.~Balachandar.
\newblock Wake structure of a transversely rotating sphere at moderate reynolds
  numbers.
\newblock \emph{J. Fluid Mech.}, 621:\penalty0 103--130, 2009.

\bibitem[Gillespie(1996)]{Gil1996}
D.~T. Gillespie.
\newblock Exact numerical simulation of the ornstein-uhlenbeck process and its
  integral.
\newblock \emph{Phys. Rev. E}, 54:\penalty0 2084--2091, 1996.

\bibitem[Jenny et~al.(2004)Jenny, Dusek, and Bouchet]{Jen2004}
M.~Jenny, J.~Dusek, and G.~Bouchet.
\newblock Instabilities and transition of a sphere falling or ascending freely
  in a newtonian fluid.
\newblock \emph{J. Fluid Mech.}, 508:\penalty0 201--239, 2004.

\bibitem[Koch and Shaqfeh(1989)]{Koc1989}
D.~L. Koch and E.~S.~G. Shaqfeh.
\newblock The instability of a dispersion of sedimenting spheroids.
\newblock \emph{J. Fluid Mech.}, 209:\penalty0 521--542, 1989.

\bibitem[Komasawa et~al.(1974)Komasawa, Kuboi, and Otake]{Kom1974}
I.~Komasawa, R.~Kuboi, and T.~Otake.
\newblock Fluid and particle motion in turbulent dispersion‰ÛÓi: Measurement of
  turbulence of liquid by continual pursuit of tracer particle motion.
\newblock \emph{Chem. Eng. Sci.}, 29\penalty0 (3):\penalty0 641 -- 650, 1974.
\newblock ISSN 0009-2509.
\newblock \doi{10.1016/0009-2509(74)80178-8}.

\bibitem[Mortensen et~al.(2007)Mortensen, Andersson, Gillissen, and
  Boersma]{Mor2007}
P.~H. Mortensen, H.~I. Andersson, J.~J.~J. Gillissen, and B.~J. Boersma.
\newblock Particle spin in a turbulent shear flow.
\newblock \emph{Phys. Fluids}, 19:\penalty0 1--4, 2007.

\bibitem[Mortensen et~al.(2008{\natexlab{a}})Mortensen, Andersson, Gillissen,
  and Boersma]{Mor2008a}
P.~H. Mortensen, H.~I. Andersson, J.~J.~J. Gillissen, and B.~J. Boersma.
\newblock Dynamics of prolate ellipsoidal particles in a turbulent channel
  flow.
\newblock \emph{Phys. Fluids}, 20:\penalty0 1--14, 2008{\natexlab{a}}.

\bibitem[Mortensen et~al.(2008{\natexlab{b}})Mortensen, Andersson, Gillissen,
  and Boersma]{Mor2008b}
P.~H. Mortensen, H.~I. Andersson, J.~J.~J. Gillissen, and B.~J. Boersma.
\newblock On the orientation of ellipsoidal particles in a turbulent shear
  flow.
\newblock \emph{Int. J. Multiphase Flow}, 34, 2008{\natexlab{b}}.

\bibitem[Norris(1998)]{Nor1998}
J.~R. Norris.
\newblock \emph{Markov Chains (Cambridge Series in Statistical and
  Probabilistic Mathematics)}.
\newblock Cambridge University Press, 1998.

\bibitem[Pope(2000)]{Pop2000}
S.~B. Pope.
\newblock \emph{Turbulent Flows}.
\newblock Cambridge University Press, 2000.

\bibitem[Shin and Koch(2005)]{Shi2005}
M.~Shin and D.~L. Koch.
\newblock Rotational and translational dispersion of fibres in isotropic
  turbulent flows.
\newblock \emph{J. Fluid Mech.}, 540, 2005.

\bibitem[Snyder and Lumley(1971)]{Sny1971}
W.~H. Snyder and J.~L. Lumley.
\newblock Some measurements of particle velocity autocorrelation functions in a
  turbulent flow.
\newblock \emph{J. Fluid Mech.}, 48:\penalty0 41--47, 1971.

\bibitem[Taylor(1921)]{Tay1921}
G.~I. Taylor.
\newblock Diffusion by continuous movements.
\newblock \emph{Proc. London Math. Soc. Ser. A}, 20:\penalty0 196--212, 1921.

\bibitem[Variano and Cowen(2008)]{Var2008}
E.~A. Variano and E.~A. Cowen.
\newblock A random-jet-stirred turbulence tank.
\newblock \emph{J. Fluid Mech.}, 604:\penalty0 1--32, 2008.

\end{thebibliography}
\bibliographystyle{plainnat}

\end{document}